\documentclass[11pt]{article}
\usepackage{amssymb,amsmath,epsfig,graphics}
\def\one{{{{\rm 1} \kern -.19em {\rm l}}}}
\def\C{{{{\rm {\mbox{\small l}}} \kern -.50em {\rm C}}}}
\def\R{{{{\rm l} \kern -.15em {\rm R}}}}
\def\N{{{{\rm l} \kern -.15em {\rm N}}}}
\def\E{{{{\rm l} \kern -.15em {\rm E}}}}
\def\P{{{{\rm l} \kern -.15em {\rm P}}}} 
\def\Z{{{{\rm Z} \kern -.35em {\rm Z}}}}
\def\1{{{{\rm 1} \kern -.35em {\rm 1}}}}

\begin{document}
\begin{sloppypar}
\vspace*{0cm}
\begin{center}
{\setlength{\baselineskip}{1.0cm}{ {\Large{\bf 
THE CONFLUENT SUPERSYMMETRY ALGORITHM \\ FOR DIRAC EQUATIONS WITH 
PSEUDOSCALAR POTENTIALS
 \\}} }}
\vspace*{1.0cm}
{\large{\sc{Alonso Contreras-Astorga}$^\dagger$} and {\sc{Axel Schulze-Halberg}$^\ddagger$}}
\indent \vspace{.3cm} 
\noindent \\	
Department of Mathematics and Actuarial Science and Department of Physics, Indiana
University Northwest, 3400 Broadway, Gary IN 46408, USA,\\ ${}^\dagger$E-mail:
aloncont@iun.edu 
\\ ${}^\dagger$E-mail:
axgeschu@iun.edu, xbataxel@gmail.com 
\end{center}

\vspace*{.5cm}
\begin{abstract}
\noindent
We introduce the confluent version of the quantum-mechanical supersymmetry (SUSY) formalism for 
the Dirac equation with a pseudoscalar potential. Application of the formalism to spectral problems is 
discussed, regularity conditions for the transformed potentials are derived, and normalizability of the 
transformed solutions is established. Our findings extend and complement former results \cite{nieto}.

\end{abstract} 

\section{Introduction}
The formalism of supersymmetry (SUSY) has become popular in quantum mechanics as a method to 
construct new solvable models. The principal scheme of SUSY 
goes back to the Darboux transformation that was first introduced in \cite{darboux} as an algorithm 
to map solutions of linear differential equations onto each other. As such, it applies to quantum-mechanical 
systems governed by equations of Schr\"odinger type. While 
the concept of Darboux transformations has been developed further and became applicable to a 
variety of linear and nonlinear equations \cite{darbbook} \cite{matveev}, it has been most successfully used 
within the SUSY framework in quantum-mechanical applications that involve spectral problems, 
see e.g. the reviews \cite{cooper} \cite{djf} and references therein. The reason for this lies in the 
fact that besides generating new systems admitting closed-form solutions, the SUSY scheme can be 
used for the modification of associated spectra. In particular, discrete spectral values corresponding 
to bound state solutions of a quantum system, can be created or removed. 
There is a large amount of literature dedicated to the study of such spectral problems, newer examples 
include applications to particular systems, involving parametric double-well potentials \cite{rosu}, 
confined harmonic oscillator intercations \cite{djoscb}, or the Swanson model \cite{sinha}. More examples 
of such applications can be found from the references in \cite{djf}. Another, 
related branch of research that involves the SUSY scheme concerns the construction of rational extensions to 
solvable potentials, such that the extended system admits solutions in terms of exceptional 
orthogonal polynomials \cite{gomez1} \cite{gomez2}, see for example the recent 
studies \cite{odake3} \cite{quesne2} \cite{quesneneu}. Roughly speaking, one can distinguish two different 
types of SUSY algorithms that apply to models governed by the Schr\"odinger equation: 
the standard algorithm that is used in the vast majority of applications, and the 
less known confluent algorithm. The difference between them manifests in the so-called auxiliary or seed 
solutions that are necessary to apply the respective algorithm. In the standard case, the latter functions 
are required to be solutions to the initial Schr\"odinger equation, while in the confluent case 
they must solve an inhomogeneous system of equations, often referred to as a Jordan chain. There is 
considerably less literature on the confluent SUSY scheme than on its standard counterpart. An introduction, 
discussion of mathematical properties and some applications can be found in \cite{baye95}, \cite{mielnik00}, \cite{djcon1}-\cite{djcon3}, as well as in 
\cite{djinv} and \cite{xbatejp}. Despite the latter references, properties of the confluent SUSY algorithm remain 
unknown, for example the link between the auxiliary solutions and regularity of the transformed potential in 
the general case. Furthermore, generalizations of the standard SUSY scheme to models not governed by 
Schr\"odinger equations, do not include the confluent case. An example for such a scenario is the Dirac 
equation. The SUSY algorithm was introduced \cite{nieto} and applied \cite{alonsod}, but only for the 
standard context. The purpose of the present work is to complement the latter results by setting up the 
confluent SUSY scheme for the Dirac equation, applicable to systems governed by pseudoscalar 
potentials. For applications of such potentials, the reader may refer to \cite{qian} \cite{decastro} 
and references therein. The remainder of this paper is organized as follows. Section 2 gives 
a brief review of the confluent SUSY formalism for 
the Schr\"odinger equation, while section 3 is devoted to the construction of the confluent 
algorithm for the Dirac equation. A discussion regarding spectral problems and regular transformed 
potentials can be found in section 4. The final section 5 presents several examples, demonstrating 
the generation of solutions to spectral problems as well as the manipulation of the corresponding spectra.

\section{The confluent SUSY scheme}
We start out by considering the following stationary Schr\"odinger equation in atomic units
\begin{eqnarray}
\Psi''+(E-U_0)~\Psi &=& 0, \label{schro}
\end{eqnarray}
where the energy $E$ is a real-valued constant and $U_0=U_0(x)$ denotes the potential. In order to apply a 
$N$-th order confluent SUSY transformation (or $N$-SUSY transformation) to a solution $\Psi$ of (\ref{schro}), we must first determine $N \geq 2$ functions 
$u_0,u_1,...,u_{N-1}$, that solve the following system of equations, 
\begin{eqnarray}
u_0''+(\lambda-U_0)~u_0 &=& 0 \label{j1} \\
u_j''+(\lambda-U_0)~u_j &=&- u_{j-1},~~~j=1,...,N-1, \label{j2}
\end{eqnarray}
introducing a real constant $\lambda$. The system (\ref{j1}), (\ref{j2}) is commonly referred to as 
Jordan chain of order $N$, while the functions $u_j$, $j=0,...,N-1$ are often called auxiliary solutions or 
transformation functions. In the conventional SUSY scheme, the auxiliary solutions must satisfy the initial Schr\"odinger 
equation at pairwise different energies \cite{bagrov}, which precisely constitutes the difference to the 
confluent algorithm that we are focusing on: here, all auxiliary solutions are associated with the same 
energy value $\lambda$. Now, once the system (\ref{j1}), (\ref{j2}) has been solved, we can construct the 
following function $\Phi$:
\begin{eqnarray}
\Phi &=& \frac{W_{u_0,...,u_{N-1},\Psi}}{W_{u_0,...,u_{N-1}}}, \label{phi}
\end{eqnarray}
where the symbol $W$ stands for the Wronskian of the auxiliary solutions in its index. The function 
$\Phi$ is a solution to the Schr\"odinger equation
\begin{eqnarray}
\Phi''+(E-U_1)~\Phi &=& 0, \label{schrot}
\end{eqnarray}
for the transformed potential $U_1$, given by the expression
\begin{eqnarray}
U_1 &=& U_0 - 2~\frac{d^2}{dx^2} \log\left(W_{u_0,...,u_{N-1}} \right)+C. \label{pottrans}
\end{eqnarray}
Here, $C$ stands for an arbitrary constant. 
While the expressions for the transformed solution (\ref{phi}) and its associated potential (\ref{pottrans}) 
look formally the same as in the conventional SUSY scheme \cite{cooper}, they are profoundly different 
due to the system (\ref{j1}), (\ref{j2}) that determines the auxiliary solutions in the confluent case. 
By means of the variation-of-constants formula, the following integral representation of the auxiliary 
solutions can be constructed \cite{xbatejp}:
\begin{eqnarray}
u_j &=& \hat{u}+u_0~ \int\limits^x \left(\int\limits^t u_0~u_{j-1}~ds \right) \frac{1}{u_0^2}~dt,~~~j=1,...,N-1, 
\label{integral}
\end{eqnarray}
where $\hat{u}$ stands for any solution of the first equation (\ref{j1}). An alternative representation for 
the second-order case $N=2$ in (\ref{j1}), (\ref{j2}) is given by \cite{bermudez}
\begin{eqnarray}
u_1 &=& \hat{u} + \frac{\partial}{\partial \lambda} ~u_0, \label{u1rep2}
\end{eqnarray}
where $\hat{u}$ stands for any solution of (\ref{j1}) and we understand $u_0$ as a function depending 
on both the variable $x$, as well as the parameter 
$\lambda$ that enters through equation (\ref{j1}) and it is a sufficiently smooth function of both arguments. It is important to point out that the particular solutions, 
given by the second terms on the right hand sides of (\ref{integral}) for $N=2$ and (\ref{u1rep2}), 
respectively, in general are not the same. The second representation (\ref{u1rep2}) is very convenient if 
the integrals in (\ref{integral}) cannot be carried out, as will be demonstrated in the application section 5.2.

\section{The confluent SUSY scheme for the Dirac equation}
In this work we focus on the following stationary, two-component Dirac equation in 
one spatial dimension \cite{nieto} \cite{peres}
\begin{eqnarray}
i~\sigma_2~\Psi' + (V_0 - E)~\Psi &=& 0, \label{dirac}
\end{eqnarray}
where $\sigma_2$ stands for the second Pauli matrix, $E$ denotes a real-valued number, and 
the spinor $\Psi=(\Psi_1,\Psi_2)^T$ represents the solution. Furthermore, we assume the potential $V_0$ 
to be of pseudoscalar form, that is, 
\begin{eqnarray}
V_0 &=& m~\sigma_3+q_0~\sigma_1, \label{pseudo}
\end{eqnarray}
for a constant, positive mass $m$, Pauli matrices $\sigma_1$, $\sigma_3$, and a function $q_0$. 
In components, the Dirac equation (\ref{dirac}) for the potential (\ref{pseudo}) can be written as
\begin{eqnarray}
\Psi_1'-q_0~\Psi_1+ \left( E+m\right) \Psi_2 &=& 0, \label{d1} \\
\Psi_2'+q_0~\Psi_2- \left( E-m \right) \Psi_1 &=& 0. \label{d2}
\end{eqnarray}
Solving the first of these equations for $\Psi_2$ gives the relation
\begin{eqnarray}
\Psi_2 &=& \frac{1}{E+m} \left(q_0~\Psi_1-\Psi_1' \right). \label{2nd}
\end{eqnarray}
Substitution of this expression into (\ref{d2}) leads to the equation
\begin{eqnarray}
\Psi_1''+\left( E^2-m^2-q_0^2-q_0'\right) \Psi_1 &=& 0. \label{schr}
\end{eqnarray}
Hence, any spinor that solves the initial Dirac equation (\ref{dirac}), is also a solution to the 
equations (\ref{2nd}), (\ref{schr}) and vice versa. Let us now rewrite (\ref{schr}) as follows
\begin{eqnarray}
\Psi_1''+\left( \epsilon-U_0 \right) \Psi_1 &=& 0, \label{sse}
\end{eqnarray}
where $\epsilon=E^2-m^2$ and $U_0 = q_0^2+q_0'$. Equation (\ref{sse}) is of Schr\"odinger form and as such 
admits the SUSY formalism. While the conventional SUSY scheme has already been discussed and applied, 
we will now focus on the confluent case and make it available to the Dirac equation. To this end, assume that 
we have found $N \geq 2$ solutions $u_0,...,u_{N-1}$ of the Jordan chain (\ref{j1}), (\ref{j2}). Then the function $\Phi_1$, given by (\ref{phi}) is a solution of 
the Schr\"odinger equation
\begin{eqnarray}
\Phi_1''+\left( \epsilon-U_1 \right) \Phi_1 &=& 0. \label{sset}
\end{eqnarray}
The potential $U_1$ is given by the expression (\ref{pottrans}), that is,
\begin{eqnarray}
U_1 &=& U_0 - 2~\frac{d^2}{dx^2} \log\left(W_{u_0,...,u_{N-1}}\right)+C, \label{U1}
\end{eqnarray}
for an arbitrary constant $C$. Now, in order to convert the transformed Schr\"odinger potential (\ref{U1}) to 
a pseudoscalar Dirac potential, we must identify a function $q_1$ that is the counterpart of $q_0$ in 
(\ref{pseudo}). Such a function $q_1$ can be obtained once the transformed Schr\"odinger equation 
(\ref{sset}) is rewritten in the form
\begin{eqnarray}
\Phi_1''+\left( E^2-m^2-q_1^2-q_1'\right) \Phi_1 &=& 0. \label{sset1}
\end{eqnarray}
Comparison of the two equations (\ref{sset}) and (\ref{sset1}) leads to the following condition for 
the sought function $q_1$:
\begin{eqnarray}
q_1^2+q_1' &=& U_1, \label{ric}
\end{eqnarray}
where $U_1$ is given in (\ref{U1}). Equation (\ref{ric}) is of Riccati type, 
the general solution of which can only be obtained from a known particular solution that will be 
determined now. First, we apply the linearizing transformation $q_1=\hat{q}'/\hat{q}$ to 
(\ref{ric}), leading to the result
\begin{eqnarray}
\hat{q}''-U_1~\hat{q} &=& 0. \label{riclin}
\end{eqnarray}
After comparison with equation (\ref{sset}) we conclude that a particular solution of (\ref{riclin}) is 
provided by a solution $\Phi_1$ of (\ref{sset}), evaluated at the energy $\epsilon=0$, that is, 
\begin{eqnarray}
\hat{q} =  \Phi_1{}_{\mid \epsilon = 0}. \label{hatq}
\end{eqnarray}
As a direct consequence we obtain a particular solution $q_{1,p}$ of the Riccati equation (\ref{ric}) as
\begin{eqnarray}
q_{1,p} ~=~ \left(\frac{\hat{q}'}{\hat{q}}\right)
~=~ \frac{d}{dx}~ \log\left( \hat{q} \right). \label{q1p}
\end{eqnarray} 
It is well-known that the general solution of our Riccati equation can be constructed by means of 
a particular solution $q_{1,p}$ via the formula
\begin{eqnarray}
q_1 &=& q_{1,p} + \frac{\exp\left(-2~\int\limits^x q_{1,p}~dt \right)}{c+\int\limits^x 
\exp\left(-2~\int\limits^t q_{1,p}~ds \right)~dt}, \label{q1pp} 
\end{eqnarray}
where $c$ denotes an arbitrary constant. Since we know that our particular solution is given in the form
(\ref{q1p}), we can incorporate the latter function into (\ref{q1pp}), leading to the simplified version
\begin{eqnarray}
q_1 &=&  \left[ \frac{d}{dx}~ \log\left(\hat{q} \right) \right] + \left(
c~\hat{q}^2 + \hat{q}^2 \int\limits^x \frac{1}{\hat{q}^2}~dt
\right)^{-1}. \label{q1}
\end{eqnarray}
Since the constant $c$ is arbitrary, expression (\ref{q1}) provides a one-parameter family of 
functions that result from the confluent SUSY transformation. In the last step it remains to convert 
(\ref{q1}) into a transformed pseudoscalar Dirac potential $V_1$, given by
\begin{eqnarray}
V_1 &=& m~\sigma_3+q_1~\sigma_1. \label{v1gen}
\end{eqnarray}
The explicit form of this potential is obtained after insertion of (\ref{q1}). We get
\begin{eqnarray}
V_1 &=& m~\sigma_3+\left\{
\left[\frac{d}{dx}~ \log\left( \hat{q} \right) \right] + \left(
c~\hat{q}+ \hat{q}^2 
\int\limits^x \frac{1}{\hat{q}}~dt
\right)^{-1}~
\right\} \sigma_1, \label{v1}
\end{eqnarray}
recall that the function $\hat{q}$ is defined in (\ref{hatq}). The one-parameter family of pseudoscalar potentials 
(\ref{v1}) enters in the transformed Dirac equation as follows
\begin{eqnarray}
i~\sigma_2~\Phi' + (V_1 - E)~\Phi &=& 0. \label{diract}
\end{eqnarray}
The first component $\Phi_1$ of the solution spinor $\Phi=(\Phi_1,\Phi_2)^T$ to our equation (\ref{diract}) 
is given in (\ref{phi}), while the second component $\Phi_2$ 
is determined through a relation analogous to (\ref{2nd}):
\begin{eqnarray}
\Phi_2 &=& \frac{1}{E+m} \left(q_1~\Phi_1-\Phi_1' \right), \label{phi2tr}
\end{eqnarray}
where the explicit form of $q_1$ can be found in (\ref{q1}). Due to the length of the expressions 
involved here, we omit to show the expanded form of $\Phi_2$.

\section{Regularity conditions and spectral problems}
While in the previous section we introduced the confluent SUSY algorithm as an entirely 
computational scheme for the Dirac equation, we will now focus on spectral problems, as they 
appear in quantum mechanics. In particular, we are interested in finding constraints, under which 
the transformed potential is free of singularities and the associated transformed solution is 
$L^2$-normalizable. Before we go into more detail, 
let us assume the Dirac equation (\ref{dirac}) to be defined on 
a real interval $D=(a,b)$, equipped with boundary conditions of Dirichlet type at the endpoints of $D$, that is, 
$\Psi(a) ~=~ \Psi(b) ~=~ (0,0)^T$, which are allowed to be understood in the sense of a limit, and the norm of a spinor given by
\begin{eqnarray}
\left|\left|\Psi \right| \right|^2 ~=~ \int_a^b \left( |\Psi_1|^2 + |\Psi_2|^2 \right)~ dx. \label{norm}
\end{eqnarray}
We further assume 
that this spectral problem admits a discrete spectrum $(E_n)$ of energies, where $n$ belongs to some 
index set, such that $\epsilon_n=E_n^2-m^2$ is uniformly bounded from below by its minimum $\epsilon_0$ 
(ground state energy).

The transformed problem, governed by equation (\ref{diract}) for the boundary conditions 
$\Phi(a) ~=~ \Phi(b) ~=~ (0,0)^T$, must have a nonsingular potential $V_1$ in order for the solutions to be 
physically meaningful. If we do not want singularities in the domain of $V_1$, we have to take care of two 
issues: to obtain a regular Schr\"odinger potential $U_1$ from the confluent SUSY transformation and to generate $q_1$ without additional singularities. 

The condition  imposed by the confluent algorithm \cite{djf} to obtain a Schr\"odinger potential $U_1$ without new singularities is that the Wronskian $W_{u_0, u_1, \dots, u_{N-1}}$ in \eqref{pottrans} must be nodeless in $(a,b)$. In other words, the Wronskian is allowed to vanish only at the points $a$ or $b$. This condition has different characteristics for each order $N$ of the SUSY transformation. For example, in the second order case $N=2$, the derivative of the Wronskian   $W_{u_0,u_1}$ can be expressed, using the Jordan chain (\ref{j1}), (\ref{j2}), as 
\begin{eqnarray}
W'_{u_0,u_1}~=~u_0 ~u_1''-u_0''~u_1~=~-u_0^2,  \nonumber 
\end{eqnarray}
which implies that for a real potential $U_0$ in \eqref{sse}, the Wronskian is a non increasing monotone function and can be written as 
\begin{eqnarray} 
W_{u_0,u_1}~=~ w_0-\int\limits^x u_0^2 ~dt, \label{wu0u1}
\end{eqnarray}
where $w_0$ is an integration constant. If we find a transformation function such that $u_0(a)=0$ or $u_0(b)=0$, then it can be always found a domain for $w_0$ to avoid zeros in the Wronskian. For the third order case $N=3$ in addition of the condition for the value in one of the boundaries of the domain $D$ it is also required that $\lambda  \leq \epsilon_0$ in \eqref{j1}, for more details of this case see \cite{djcon3}. 
 
Now, the regularity of the pseudoscalar Dirac potential (\ref{v1}) also depends  on the function $q_1$, as given in (\ref{q1p}) and (\ref{q1}).  Let us consider the 
particular case (\ref{q1p}), our condition is to find a solution $\hat{q}=\Phi_1{}_{\mid \epsilon = 0}$ that 
is free of zeros in $(a,b)$. It 
generates a nonsingular parametrizing function $q_1$ if we choose the latter function as the 
particular solution $\hat{q}$ in (\ref{q1p}). In case we use the general solution (\ref{q1}) for $q_1$, rather than 
(\ref{q1p}), we can avoid singularities in $(a,b)$ by additionally imposing that $\int^x 1/\hat{q}^2 ~dt$ is 
bounded from above or below. This choice implies that there is a constant $c$ in (\ref{q1}), such that 
the denominator in the second term does not vanish.

When the confluent SUSY algorithm is applied to the Schr\"odinger equation some properties are conserved, in particular the spectral properties which reminds almost equal, the spectral manipulation that can be done are: to add an eigenvalue, to delete one or to obtain a new equation with the same spectrum. The position of this possible extra eigenvalue is in general arbitrary but every order $N$ of the transformation has its own peculiarities. To observe if one extra eigenvalue is inserted or suppressed in the spectrum of the transformed equation special attention have to be paid when $\epsilon = \lambda$. Notice that $u_0$ is a solution of \eqref{schro} and the corresponding transformed solution \eqref{phi} is the trivial. A linear independent solution $v_0$ can be obtained asking $W_{u_0, v_0}=1$ \cite{kamke}, that is 
\begin{eqnarray}
v_0 &=& u_0 \int\limits^x \frac{1}{u_0^2}~dt, \label{v0ex2}
\end{eqnarray}
a direct substitution demonstrate that $v_0$ is solution of \eqref{j1}.  Now, if we transform this solution using the rule \eqref{phi}, then   
\begin{eqnarray}
\Phi_1 &=& \frac{W_{u_0, \dots, u_{N-1},v_0}}{W_{u_0, \dots, u_{N-1}}} ~=~  \frac{W_{u_0, \dots, u_{N-2}}}{W_{u_0, \dots, u_{N-1}}}. \label{missing state}
\end{eqnarray}
To obtain the last equality first the columns of the Wronskian determinant were interchanged, 
\begin{eqnarray}
W_{u_0, \dots, u_{N-1},v_0}&=& (-1)^{N-1}W_{u_0,v_0, u_1, \dots, u_{N-1}}, \nonumber
\end{eqnarray} 
then all the second and higher derivatives of the transformation functions were replaced by the transformation functions themselves and their first derivative using the Jordan chain \eqref{j1}, \eqref{j2}. With elementary row operations an upper triangular block determinant can be obtained,
\begin{eqnarray}
W_{u_0, \dots, u_{N-1},v_0}&=& \begin{vmatrix}
u_0& v_0 & u_1 & \dots & u_{N-1} \\
u'_0& v'_0 & u'_1 & \dots & u'_{N-1} \\
0 & 0 & u_0 & \dots & u_{N-2} \\
0 & 0 &u'_0 & \dots & u'_{N-2} \\
\vdots & \vdots &  & \ddots &  \\
0 & 0 &u^{(N-3)}_0 & \dots & u^{(N-3)}_{N-2}\\
\end{vmatrix} ~=~ W_{u_0, v_0} W_{u_0, u_1, \dots, u_{N-2}} \nonumber
\end{eqnarray}
and since by construction $W_{u_0, v_0}=1$, the equation \eqref{missing state} has been demonstrated. If the solution \eqref{missing state} is $L^2-$normalizable, then $\lambda$ is an element of the spectrum of the problem \eqref{schrot} with Dirichlet boundary conditions $\Phi_1(a)=\Phi_1(b)=0$. Note again the importance of study the zeros of $W_{u_0, \dots, u_{N-1}}$, in this case to know if a element of the spectrum is added or deleted.

The confluent SUSY algorithm for the Dirac equation has an extra liberty in the spectral design point of view. The presence of the constant $C$ in  \eqref{U1} has no importance for the Schr\"odinger equation because it can be interpreted as a shift of the zero energy point, which is arbitrary. If $\epsilon_n$ is an eigenvalue of \eqref{sse} then $\epsilon_n + C$ will be of \eqref{sset}, in the situation that it is not suppressed by the confluent SUSY algorithm. As a result The corresponding eigenvalue of the transformed Dirac equation given in \eqref{diract} will be $|E_n| = \sqrt{\epsilon_n + C + m^2}$. The constant $C$ will shift every element of the spectrum in a non-linear fashion, and also the gap between the positive and negative elements of the spectrum can be modulated.

\section{Applications}
We will now present three applications of the confluent SUSY algorithm for the Dirac equation, involving 
potentials of physical interest. In the first application we study a Coulomb-like potential, defined on the 
positive semiaxis. The second application is devoted to a linear potential plus a rational extension that 
is defined on the whole real line and features a link to the harmonic oscillator. In the last application we 
consider a trigonometric interaction on a bounded interval.

\subsection{Coulomb-type system}
\begin{figure}[h]
\begin{center}
\epsfig{file=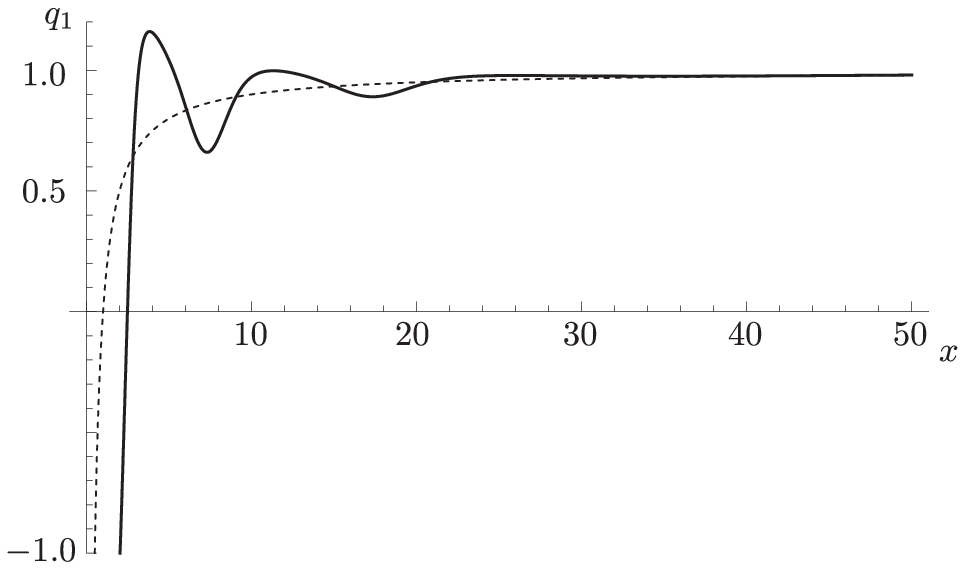,width=7.5cm} \hspace{.4cm}
\epsfig{file=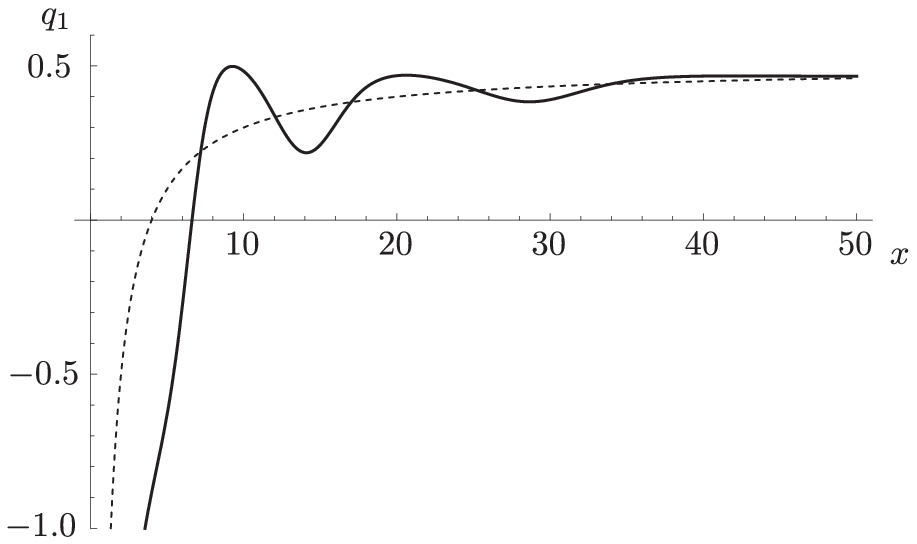,width=7.5cm}
\caption{the functions $q_0$ (dotted curve) and $q_1$ (solid curve). In order to generate $q_1$, we removed 
the spectral values $\pm |E_{2,1}|$ in the left figure, and $\pm |E_{2,2}|$ in the right figure.} 
\label{hydrogen}
\end{center}
\end{figure}

Let us consider the following boundary-value problem of Dirichlet type for the Dirac equation
\begin{eqnarray}
i~\sigma_2~\Psi' + (V_0 - E)~\Psi &=& 0,~~~~x \in \left(0,\infty \right) \label{bvpx1} \\[1ex]
\Psi(0) ~=~ \lim\limits_{x \rightarrow \infty} \Psi(x) &=& (0,0)^T, \label{bvpx2}
\end{eqnarray}
where the pseudoscalar potential $V_0$ is given by
\begin{eqnarray}
V_0 &=& m~\sigma_3+ \left(\frac{1}{\ell}-\frac{\ell}{x} \right)\sigma_1, \label{pseudoex1}
\end{eqnarray}
for a natural number $\ell$. Comparison of (\ref{pseudoex1}) with the general form 
(\ref{pseudo}) shows that the parametrizing function $q_0$ of the Dirac potential reads
\begin{eqnarray}
q_0 &=& \frac{1}{\ell}-\frac{\ell}{x}. \label{q0ex1}
\end{eqnarray}
The boundary-value problem (\ref{bvpx1}), (\ref{bvpx2}) admits a discrete spectrum of energy values 
$E_{n,\ell}$, where $n$ is a nonnegative integer:
\begin{eqnarray}
|E_{n,\ell}| &=& \sqrt{m^2+ \frac{1}{\ell^2}-\frac{1}{(1+\ell+n)^2}}. \label{ecoul}
\end{eqnarray}
The corresponding solutions $\Psi=(\Psi_1, \Psi_2)^T$ of our problem (\ref{bvpx1}), (\ref{bvpx2}) have 
the following explicit form
\begin{eqnarray}
\Psi_1 &=&  x^{\ell+1}~ \exp\left(-\frac{x}{n+\ell+1}\right) L_n^{2\ell+1}\left(  \frac{2x}{n+\ell+1}\right) 
\label{psi1ex1} 
\\[1ex] 
\Psi_2 &=& \frac{x^\ell}{E_{n,\ell}+m}~\exp\left(-\frac{x}{n+\ell+1}\right)
\left\{\frac{2~x}{n+\ell+1}~L_{n-1}^{2 \ell+2}\left(  \frac{2~x}{n+\ell+1}\right)
+ \right. \nonumber \\[1ex]
&+& \left. \left[-2~\ell-1+\frac{(n+2~\ell+1)~x}{\ell~(n+\ell+1)} \right]
L_n^{2\ell+1}\left(  \frac{2x}{n+\ell+1}\right)
\right\}, \nonumber
\end{eqnarray}
where $L$ stands for a generalized Laguerre function \cite{abram} and $E_{n,\ell}$ is defined in 
(\ref{ecoul}). Before we continue with the construction of our SUSY transformation, let us briefly 
recall that the first component (\ref{psi1ex1}) of the Dirac solution also solves the Schr\"odinger 
equation (\ref{sse}) for the potential $U_0$, which in the present case is given by 
\begin{eqnarray}
U_0 &=& q_0^2+q_0' ~=~ \frac{\ell (\ell+1)}{x^2}-\frac{2}{x}+\frac{1}{\ell^2}. \label{potential coulomb}
\end{eqnarray}
This is a shifted Coulomb potential with centrifugal barrier, which is well-known to admit a discrete spectrum, 
confirming (\ref{ecoul}). We will now apply the confluent 2-SUSY algorithm to the 
boundary-value problem (\ref{bvpx1}), (\ref{bvpx2}), such that two values are removed from the discrete 
spectrum of the transformed problem. More precisely, we are aiming at removing a pair of spectral 
values $\pm |E_{n_0, \ell}|$, where $n_0$ is a nonnegative integer. In order to perform our transformation, we 
need two transformation functions $u_0$ and $u_1$, together with an associated constant $\lambda$, 
that solve the Jordan chain (\ref{j1}), (\ref{j2}) for $N=2$. We choose the first of these functions as 
$u_0=\Psi_1{}_{\mid n = n_0}$, which solves (\ref{j1}) for the constant $\lambda=E_{n_0,\ell}^2 - m^2$, see 
(\ref{ecoul}). We will comment below on the consequences of choosing $\lambda$ as we did. 
Now, the second transformation function $u_1$ we generate by means of the integral representation 
(\ref{integral}) for $\hat{u}=0$, that is,
\begin{eqnarray}
u_1 &=& u_0~ \int\limits^x \left(\int\limits^t u_0^2~ds \right) \frac{1}{u_0^2}~dt. \label{aux2ex1}
\end{eqnarray}
This integral can be evaluated in closed form, but due to the length of the resulting expression we omit to show 
the explicit form of $u_1$ here. Now that we have the two required transformation functions, let us use the 
conditions for regularity and normalizability of the transformed potential and solution, respectively. These 
conditions, developed in section 4, state that the Wronskian (\ref{wu0u1}) 
of the transformation functions must be nonzero inside the domain $D$ of our boundary-value problem, 
given by $D=(0,\infty)$. According to (\ref{wu0u1}), we have the following Wronskian
\begin{eqnarray}
W_{u_0,u_1} ~=~ w_0-\int\limits^x u_0^2 ~dt ~=~ w_0-
\int\limits^x  t^{2 \ell+2}~ \exp\left(-\frac{2 t}{n_0+\ell+1}\right) 
\left[L_{n_0}^{2\ell+1}\left(  \frac{2t}{n_0+\ell+1}\right) \right]^2 dt. \label{wex1}
\end{eqnarray}
Since the integrand is positive, its integral is a strictly increasing function, which due to its continuity 
can have at most one zero. If we choose the integration constant $w_0$ as 
\begin{eqnarray}
w_0 &=& \left(\int\limits^x u_0^2 ~dt\right)_{\Big| x=0}, \label{wex11}
\end{eqnarray}
then we know that (\ref{wex1}) has exactly one zero, which is located at $x=0$. We are therefore guaranteed 
that the function (\ref{U1}) is free of singularities inside its domain $D=(0,\infty)$. Next, let us comment on 
the choice of $\lambda=E_{n_0,\ell}^2 - m^2$, which determines integrability of the transformed solution $\Phi_1$ 
to equation (\ref{sset}). Since by the election $\lambda$ matches one of the spectral values admitted by the boundary-value 
problem (\ref{schrot}), for the potential  (\ref{potential coulomb}) and $\Psi_1(0)=\Psi_1(\infty)=0$, and also because the Wronskian $W_{u_0,u_1}$ vanishes at the origin, the corresponding solution \eqref{missing state} will not be $L^2-$nomalizable 
and as a consequence the element
\begin{eqnarray}
\epsilon_{n_0} &=& E_{n_0}^2-m^2 ~=~ \frac{1}{\ell^2}-\frac{1}{(1+\ell+n_0)^2} \nonumber
\end{eqnarray}
will be removed from the transformed Schr\"odinger problem with the same boundary condition and using $C=0$. At the Dirac level, this corresponds to the pair of numbers $\pm |E_{n_0,\ell}|$ that is missing in the discrete 
spectrum of the transformed problem. In order to set up the potential associated with the latter problem, we 
must construct the parametrizing function $q_1$, using either (\ref{q1p}) or (\ref{q1}). For the sake of 
simplicity, we will take the first option, that is,
\begin{eqnarray}
q_1 &=& \frac{d}{dx}~\log\left(\Phi_1{}_{\mid \epsilon=0}\right)~=~ \frac{d}{dx}~\log\left(\Phi_1{}_{\mid n=-1}\right). 
\label{q1ex1}
\end{eqnarray} 
Unfortunately, the explicit form of this function is very long, such that we do not show it here. The same is true 
for the transformed solution $\Phi=(\Phi_1,\Phi_2)^T$, which is constructed according to (\ref{phi}) and 
(\ref{phi2tr}). Recall that the transformation functions are given by $u_0=\Psi_1{}_{\mid n = n_0}$ and 
(\ref{aux2ex1}),
\begin{eqnarray}
\Phi_1 &=& \frac{W_{u_0,u_1,\Psi_1}}{W_{u_0,u_1}} \label{phi1ex1} \\[1ex]
\Phi_2 &=&  \frac{1}{E_{n,\ell}+m} \left(q_1~\Phi_1-\Phi_1' \right) ~=~ 
\frac{1}{E_{n,\ell}+m} \left\{ \left[\frac{d}{dx}~\log\left(\Phi_1{}_{\mid n=-1}\right)\right] \Phi_1-\Phi_1' \right\} 
\label{phi2ex1}
\end{eqnarray}
where $q_1$ is determined by (\ref{q1ex1}). The functions (\ref{phi1ex1}), (\ref{phi2ex1}) provide a solution 
to the transformed boundary-value problem, given by 
\begin{eqnarray}
i~\sigma_2~\Phi' + (V_1 - E_{n,\ell})~\Phi &=& 0,~~~~x \in \left(0,\infty \right) \label{bvpx1t} \\[1ex]
\Phi(0) ~=~ \lim\limits_{x \rightarrow \infty} \Phi(x) &=& (0,0)^T, \label{bvpx2t}
\end{eqnarray}
for a pseudoscalar potential $V_1$ that is obtained by inserting the parametrizing function (\ref{q1ex1}) into 
the general form (\ref{v1gen}):
\begin{eqnarray}
V_1 &=& m~\sigma_3+ \left[\frac{d}{dx}~\log\left(\Phi_1{}_{\mid n=-1}\right) \right]\sigma_1. \nonumber
\end{eqnarray}
The discrete spectrum of the transformed boundary-value problem (\ref{bvpx1t}), (\ref{bvpx2t}) is given 
by (\ref{ecoul}) except for the two values that correspond to $n=n_0$. Before we complete this example, 
let us visualize special cases of the initial and transformed parametrizing functions $q_0$ and $q_1$, as 
given in (\ref{q0ex1}) and (\ref{q1ex1}), respectively, in figure \ref{hydrogen}. The figure reveals that 
the functions $q_1$ diverge stronger at zero than their counterparts $q_0$, which can be understood by 
analyzing the transformation and by determining the behaviour of the transformed parametrizing function 
$q_1$ close to $x=0$. To this end, let us recall that the transformation function $u_0$ is defined by 
$u_0=\Psi_1{}_{\mid n = n_0}$, where $\Psi_1$ can be found in (\ref{psi1ex1}). Since close to $x=0$ we have $u_0 \propto x^{\ell+1}$, the Wronskian (\ref{wex1}), (\ref{wex11}) 
shows the following asymptotic behaviour
\begin{eqnarray}
W_{u_0,u_1} &\propto& \int\limits^x t^{2(\ell+1)}~dt ~=~ x^{2\ell+3}. \nonumber
\end{eqnarray} 
Let us express this Wronskian in the convenient way
\begin{eqnarray}
W_{u_0,u_1}=x^{2 \ell+3}~ \mathcal{W}, \label{ww}
\end{eqnarray}
where the function $\mathcal{W}$ catches the behaviour of the Wronskian away from $x=0$, and as such 
is bounded there. Substitution of 
the expression (\ref{ww}) into the general form (\ref{U1}) for the transformed potential function $U_1$, the 
latter function renders in the form
\begin{eqnarray}
U_1=\frac{(\ell+2)(\ell+3)}{x^2}-\frac{2}{x}-2~\frac{d^2}{dx^2}~\log\left({\mathcal{W}} \right), \label{wwpot}
\end{eqnarray}
Since we would like to determine the behavior of the transformed function $q_1$ in (\ref{q1ex1}), it is important to study the possible behavior of the solutions of (\ref{sset}) for the potential (\ref{wwpot}). This equation allows two types of solutions according to their behavior close to the origin: the first vanishes in the origin as $\Phi_1{}_{\mid n = -1} \propto x^{\ell +3}$ and the second diverges as $\Phi_1{}_{\mid n = -1} \propto x^{-(\ell +2)}$, in this example we used the latter and as a consequence $\Phi_1{}_{\mid n = -1}$ can then be written as  
\begin{eqnarray}
\Phi_1{}_{\mid n = -1} &=& x^{-(\ell+2)}~ \varphi, \nonumber
\end{eqnarray}
where the function $\varphi$ is bounded at $x=0$. The latter expression for $\Phi_1{}_{\mid n = -1}$ can now 
be plugged into (\ref{q1ex1}), yielding the constraint
\begin{eqnarray}
q_1=-\frac{\ell+2}{x}+\frac{\varphi'}{\varphi}, \label{q1sing}
\end{eqnarray}
note that the term $\varphi'/ \varphi$ is bounded at $x=0$.
On comparing (\ref{q1sing}) to (\ref{q0ex1}), we observe that the coefficient of the $-x^{-1}$ term is greater for 
$q_1$, which explains why the transformed function $q_1$ diverges faster than $q_0$ at $x=0$.

\subsection{Harmonic oscillator system}
\begin{figure}[h]
\begin{center}
\epsfig{file=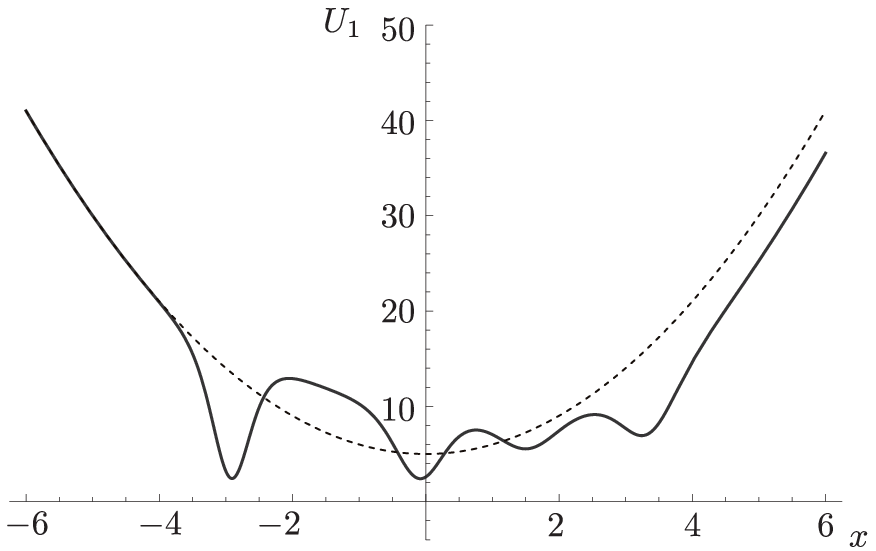,width=8cm}
\caption{The potentials $U_0=x^2+5$ (dotted curve) and its counterpart $U_1$ (solid curve), 
as given in (\ref{u1ex2}) for $\lambda=9.1$ and $B=-0.01$.} 
\label{oscillador Schrodinger}
\end{center}
\end{figure}

\begin{figure}[h]
\begin{center}
\epsfig{file=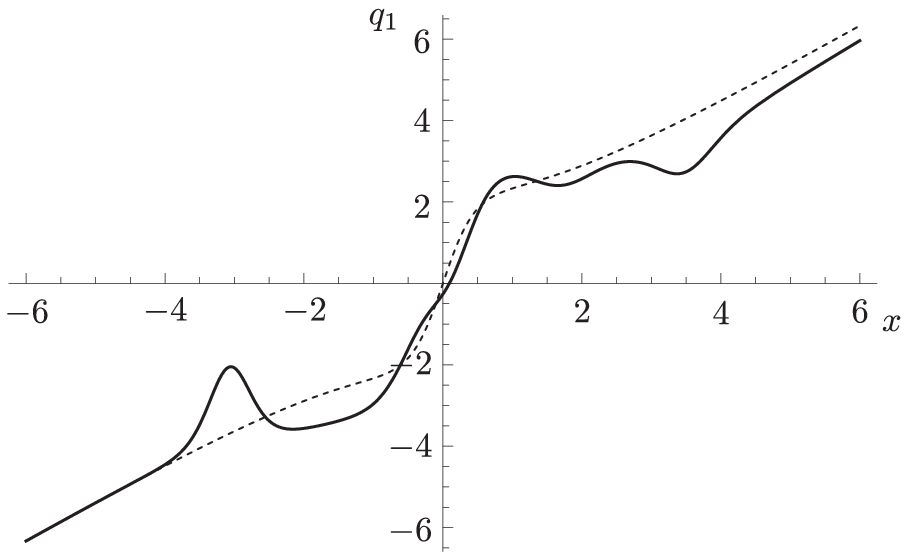,width=7.5cm} \hspace{.4cm}
\epsfig{file=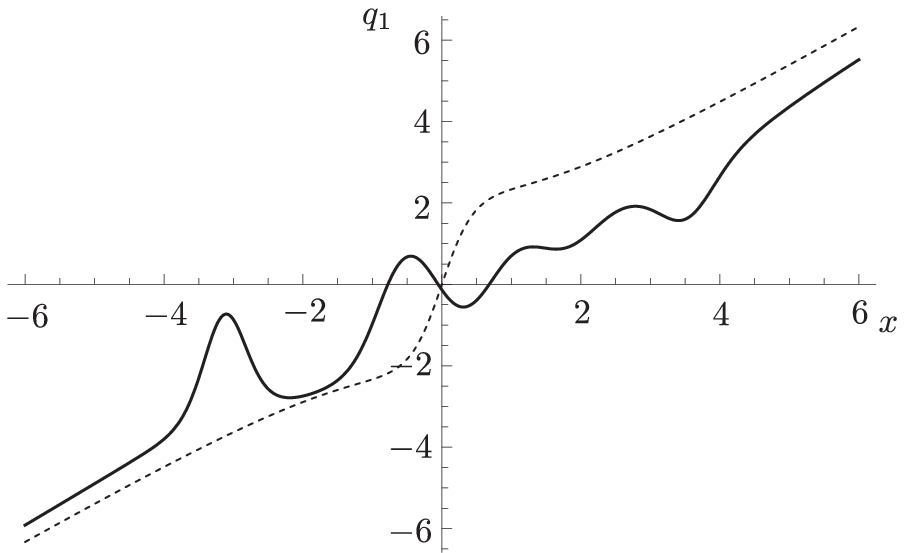,width=7.5cm}
\caption{The initial and transformed parametrizing functions $q_0$ (dotted curve) and $q_1$ (solid curve), as given in 
(\ref{5}) and (\ref{q1finalex2}), respectively. The settings are $\lambda=9.1$, $B=-0.01$, $C=0$ (left plot) and 
$\lambda=4.1$, $B=-0.01$, $C=-5$ in (\ref{u1ex2}) (right plot).
} 
\label{oscillator Dirac}
\end{center}
\end{figure}

In contrast to the previous example, this time we will start out from a boundary-value problem for 
the Schr\"odinger equation (\ref{schr}), and generate the Dirac SUSY partners arising from the 
confluent algorithm. Let us consider the following problem
\begin{eqnarray}
\Psi_1''+\left( \epsilon-x^2+A \right) \Psi_1 &=& 0,~~~ x \in (-\infty,\infty) \label{bvpschr1} \\[1ex]
\lim\limits_{|x| \rightarrow \infty} \Psi_1(x) &=& 0, \label{bvpschr2}
\end{eqnarray}
where the real constant $A$ is a free parameter that will prove useful below. By inspection of (\ref{schr}) we 
observe that in the present case
\begin{eqnarray}
U_0 &=& x^2-A, \label{u0ex2}
\end{eqnarray}
which is a shifted harmonic oscillator interaction. The boundary-value problem 
(\ref{bvpschr1}), (\ref{bvpschr2}) admits an infinite discrete spectrum 
$(\epsilon_n)$, where $n$ is a nonnegative integer:
\begin{eqnarray}
\epsilon_n &=& 2~n+1-A. \label{epsex2}
\end{eqnarray}
The associated, $L^2$-normalizable solutions are given by
\begin{eqnarray}
\Psi_1 &=& \exp\left(-\frac{x^2}{2} \right) H_n(x), \label{psi1ex2}
\end{eqnarray}
where $H$ denotes a Hermite polynomial \cite{abram}. Before we continue, let us state the general solution 
$u$ of our Schr\"odinger equation (\ref{bvpschr1}), which will prove useful when we construct the Dirac problem.
\begin{eqnarray}
u = c_1~\exp\left(-\frac{x^2}{2} \right) {}_1F_1\left(\frac{1-\epsilon-A}{4},\frac{1}{2},x^2 \right)+
c_2~\exp\left(-\frac{x^2}{2} \right) x~{}_1F_1\left(\frac{3-\epsilon-A}{4},\frac{3}{2},x^2 \right), \label{general}
\end{eqnarray}
for two free constants $c_1$ and $c_2$. Furthermore, the symbol ${}_1F_1$ stands for the 
confluent hypergeometric function \cite{abram}. We will now set up a boundary-value problem for the Dirac equation (\ref{dirac}) that can be related to 
(\ref{bvpschr1}), (\ref{bvpschr2}). To this end, we determine the parametrizing function $q_0$ of the 
pseudoscalar Dirac potential to be constructed. According to $U_0 = q_0^2+q_0'$ and (\ref{u0ex2}), 
we must solve the equation
\begin{eqnarray}
x^2-A &=& q_0^2+q_0', \nonumber
\end{eqnarray}
with respect to $q_0$. Since this equation has the same form as its counterpart (\ref{ric}), we can use the 
same solution method. To keep calculations simple, we will restrict ourselves to the particular solution 
inspired by (\ref{q1p}). We therefore need to evaluate a solution of the Schr\"odinger equation (\ref{bvpschr1}) 
at zero energy, that is, $\epsilon=0$. After setting $\epsilon=0$ in the general solution (\ref{general}) we obtain
\begin{eqnarray}
q_0 &=&  \frac{d}{dx}~\log\left(u_{\mid \epsilon = 0}\right) \nonumber \\[1ex]
&=& \frac{d}{dx}~\log\left[
c_1~\exp\left(-\frac{x^2}{2} \right) {}_1F_1\left(\frac{1-A}{4},\frac{1}{2},x^2 \right)+
c_2~\exp\left(-\frac{x^2}{2} \right) x~{}_1F_1\left(\frac{3-A}{4},\frac{3}{2},x^2 \right)\right]. \nonumber
\nonumber 
\end{eqnarray}
While in general this expression consists of intricate special functions, it can be made rational and real-valued 
by appropriate choices of the constants $c_1, c_2$ and $A$. In particular, one can obtain rational functions 
$q_0$ if the constant $A$ is chosen as $A=-4k-1$ for a nonnegative integer $k$ 
and if $c_2=0$, as the following examples show:
\begin{align}
A~=~-1,~c_2~=~0 \qquad  \Rightarrow \qquad &q_0~=~x \nonumber \\
A~=~-5,~c_2~=~0 \qquad  \Rightarrow \qquad &q_0~=~x+\frac{4~x}{2~x^2+1} \label{5} \\
A~=~-9,~c_2~=~0 \qquad  \Rightarrow \qquad &q_0~=~x+\frac{8~(2~x^3-3~x)}{4~x^4+12~x^2+3}, \nonumber 
\end{align}
Let us pick the case (\ref{5}) for our parametrizing function $q_0$. This choice determines a 
pseudoscalar potential (\ref{pseudo}) of the form
\begin{eqnarray}
V_0 &=& m~\sigma_3+ \left(x+\frac{4~x}{2~x^2+1} \right)\sigma_1, \nonumber
\end{eqnarray}
which enters in the corresponding boundary-value problem for the Dirac equation that reads
\begin{eqnarray}
i~\sigma_2~\Psi' + (V_0 - E)~\Psi &=& 0,~~~~x \in \left(-\infty,\infty \right) \nonumber \\[1ex]
\lim\limits_{x \rightarrow -\infty} \Psi(x) ~=~ \lim\limits_{x \rightarrow \infty} \Psi(x) &=& (0,0)^T. \nonumber
\end{eqnarray}
The discrete spectrum admitted by this problem can be found by plugging (\ref{epsex2}) into 
the equation $\epsilon=E^2-m^2$ and solving for $E$:
\begin{eqnarray}
|E_n| &=& \sqrt{2~n+6+m^2}. \label{eex2}
\end{eqnarray}
The first component $\Psi_1$ of the corresponding solution $\Psi=(\Psi_1,\Psi_2)$ to this problem is given in (\ref{psi1ex2}), while 
the second component $\Psi_2$ can be generated by means of (\ref{2nd}), that is,
\begin{eqnarray}
\Psi_1 \hspace{-.2cm} &=& \hspace{-.2cm} \exp\left(-\frac{x^2}{2} \right) H_n(x) \nonumber \\[1ex]
\Psi_2 \hspace{-.2cm} &=& \hspace{-.2cm} \frac{1}{\sqrt{2~n+6+m^2}+m}
\left\{
\left(x+\frac{4~x}{2~x^2+1}\right) \exp\left(-\frac{x^2}{2} \right) H_n(x)-\frac{d}{dx} 
\left[ 
\exp\left(-\frac{x^2}{2} \right) H_n(x)
\right]
\right\}. \nonumber
\end{eqnarray}
We are now ready to apply a confluent 2-SUSY transformation to the Schr\"odinger problem 
(\ref{bvpschr1}), (\ref{bvpschr2}). To this end, we need two transformation functions $u_0,u_1$ that 
solve the system (\ref{j1}), (\ref{j2}). Let us choose the function $u_0$ as a special case of the general 
solution (\ref{general}) for the settings
\begin{eqnarray}
\epsilon ~=~ \lambda \qquad \qquad \qquad c_1 ~=~ 1 \qquad \qquad \qquad c_2 ~=~ \frac{2~\Gamma\left(2-\frac{\lambda}{4}\right)}
{\Gamma\left(\frac{3}{2}-\frac{\lambda}{4}\right)}. \label{u0set}
\end{eqnarray}
A particular effect of this choice for $c_1$ and $c_2$ is that the function $u_0$ vanishes at negative infinity, 
while it becomes unbounded at positive infinity. This behavior is one of the condition needed to obtain a Wronskian free of zeros. It remains to determine a 
second transformation function $u_1$ in order to solve the system (\ref{j1}), (\ref{j2}). In the present case 
the integral representation (\ref{integral}) for $j=1$ does not yield the sought function $u_1$ in closed form, 
because the integrals cannot be evaluated, given that $u_0$ is given by (\ref{general}) and (\ref{u0set}) for a general value of $\lambda$. 
Therefore, in order to avoid integration, we will resort to the differential formula (\ref{u1rep2}), which 
can be written in the form
\begin{eqnarray}
u_1 &=& \hat{u} + \frac{\partial}{\partial \lambda}~u_0 ~=~u_0+
B~v_0 +\frac{\partial}{\partial \lambda}~u_0, \label{u1new}
\end{eqnarray}
for an arbitrary real constant $B$ and a function $v_0$ defined by \eqref{v0ex2}. Remind that $u_0$ and $v_0$ are linearly independent solutions of equation (\ref{j1}) with Wronskian 
$W_{u_0,v_0}=1$. The useful expression for the derivative with respect the first parameter of the hypergeometric function can be found in \cite{ancarani08}. Let us point out that 
we will not have to resolve the integral that was introduced in (\ref{u1new}) due to our 
knowledge about the Wronskian of $u_0$ and $v_0$. Now that we have determined the latter two 
transformation functions through (\ref{general}), (\ref{u0set}), and (\ref{u1new}), we can perform a 
confluent 2-SUSY transformation on the boundary-value problem (\ref{bvpschr1}), (\ref{bvpschr2}), 
recall that we chose $A=-5$. According to (\ref{U1}) for $N=2$, the potential $U_1$ in the transformed 
Schr\"odinger equation (\ref{sset}) is given by
\begin{eqnarray}
U_1 &=& U_0  - 2~\frac{d^2}{dx^2}~ \log\left(W_{u_0,u_1}\right) 
~=~ x^2+5 - 2~\frac{d^2}{dx^2}~ \log\left(W_{u_0,\frac{\partial u_0}{\partial \lambda} }+B\right). \label{u1ex2}
\end{eqnarray}
where the constant $C$ was set to zero. Observe that the Wronskian $W_{u_0,u_1}$ in the logarithm's 
argument can be simplified by substituting the explicit form (\ref{u1new}) of $u_1$ and then expanding the 
determinant \cite{bermudez}. As a result, the integral contained in (\ref{u1new}) does not appear in the 
Wronskian anymore. For the sake of brevity we do not show the explicit form of the transformed 
potential (\ref{u1ex2}). The transformed solution $\Phi_1$ of the equation (\ref{sset}) is constructed 
by means of (\ref{phi}) for $N=2$, that is,
\begin{eqnarray}
\Phi_1 &=& \frac{W_{u_0,u_1,\Psi_1}}{W_{u_0,u_1}} ~=~ \Psi_1'' + \frac{u_0^2}{W_{u_0,\frac{\partial u_0}{\partial \lambda} }+B} \Psi_1 - \left[ U_0 - \lambda + \frac{u_0 u_0'}{W_{u_0,\frac{\partial u_0}{\partial \lambda} }+B} \right] \Psi_1. \label{phi1ex2}
\end{eqnarray}
Note that the Wronskian in the numerator was obtained by Laplace expansion, see \cite{bermudez} 
for details. As in the case of (\ref{u1ex2}), the integral in (\ref{u1new}) does not appear, such that we do not 
need to resolve it. The $L^2$-normalizable functions (\ref{phi1ex2}) satisfy the spectral problem governed by 
(\ref{sset}) and (\ref{u1ex2}), subject to the boundary conditions implying 
that $\Phi_1$ must vanish at the infinities. We omit to substitute the explicit forms of the functions involved 
in (\ref{phi1ex2}) due to their length. Since we chose the parameters (\ref{u0set}), the discrete 
spectrum of the transformed problem is given by (\ref{epsex2}), and contains an additional 
spectral value $\lambda$. The associated 
$L^2$-normalizable solution can be found substituting $u_0$ with the parameters \eqref{u0set} and \eqref{u1new} in \eqref{missing state}. This yields
\begin{eqnarray}
\Phi_1 &=& \frac{W_{u_0,u_1,v_0}}{W_{u_0,u_1}} ~=~ 
\frac{u_0}{ W_{u_0,\frac{\partial u_0 }{\partial \lambda}}+B}, \label{missing}
\end{eqnarray}
where the Wronskian in the numerator was simplified similarly to its counterpart in (\ref{phi1ex2}). Again, for 
reasons of brevity, we do not include the full form of (\ref{missing}). In figure 
\ref{oscillador Schrodinger} we show a plot of typical potential functions $U_0$ and $U_1$, as given in 
(\ref{u0ex2}) for $A=-5$ and (\ref{u1ex2}), respectively.
We are now ready to generate a Dirac boundary-value problem from the transformed 
Schr\"odinger equation (\ref{sset}) with potential (\ref{u1ex2}). To this end, we must generate the 
parametrizing function $q_1$ that will enter in the pseudoscalar potential of the Dirac equation. 
According to (\ref{q1p}), in the simplest case we can construct $q_1$ through a solution of (\ref{sset}), 
(\ref{u1ex2}) that is taken at $\epsilon=0$. If this solution does not have any nodes, then the function 
$q_1$ will be regular on the whole real line. Let us construct $q_1$ as follows, using our transformation 
functions $u_0$ and $u_1$ as in (\ref{phi1ex2}):
\begin{eqnarray}
q_1 &=& \frac{d}{dx} ~\log\left(\frac{W_{u_0,u_1,u_{\mid \epsilon=c_2=0}}}{W_{u_0,u_1}} \right) ~=~ 
\frac{d}{dx} ~\log\left(\frac{W_{u_0,u_1,u_{\mid \epsilon=c_2=0}}}
{W_{u_0, \frac{\partial u_0 }{\partial \lambda}}+B} \right), \label{q1finalex2}
\end{eqnarray}
where $u$ is the general solution (\ref{general}) that is taken here at $\epsilon=c_2=0$. Now that we have 
determined the parametrizing function $q_1$ for the pseudoscalar potential of the transformed Dirac equation, 
it remains to construct the second component $\Phi_2$ of its solution $\Phi=(\Phi_1,\Phi_2)^T$. 
According to (\ref{phi2tr}), we have
\begin{eqnarray}
\Phi_2 &=& \frac{1}{E+m} \left(q_1~\Phi_1-\Phi_1' \right). \label{phi2ex2}
\end{eqnarray}
In this expression $q_1$ must be substituted from (\ref{q1finalex2}), while the function $\Phi_1$ can be 
either taken from the solution set (\ref{phi1ex2}) or it is given by the single solution (\ref{missing}). The 
constant $E$ are elements of the discrete spectrum that is determined from (\ref{eex2}) as follows
\begin{eqnarray}
(E_n) &=& \left\{\pm \sqrt{2~n+6+m^2}: n=0,1,2,3...\right\} \cup \left\{\pm \sqrt{\lambda+m^2} \right\}, \nonumber
\end{eqnarray}
where the last contribution contains the newly created spectral value $\lambda$ associated with (\ref{missing}). 
The functions (\ref{phi1ex2}), (\ref{phi2ex2}) solve the transformed boundary-value problem
\begin{eqnarray}
i~\sigma_2~\Phi' + (V_1 - E)~\Phi &=& 0,~~~~x \in \left(0,\infty \right) \label{bvpx1t2} \\[1ex]
\Phi(0) ~=~ \lim\limits_{x \rightarrow \infty} \Phi(x) &=& (0,0)^T, \label{bvpx2t2}
\end{eqnarray}
the pseudoscalar potential of which is given by (\ref{v1}), that is,
\begin{eqnarray}
V_1 &=& m~\sigma_3+ \left[\frac{d}{dx} ~\log\left(\frac{W_{u_0,u_1,u_{\mid \epsilon=c_2=0}}}
{W_{u_0, \frac{\partial u_0 }{\partial \lambda}}+B} \right)\right]\sigma_1. \nonumber
\end{eqnarray}
Figure \ref{oscillator Dirac} shows two plots of the parametrizing functions $q_0$ and $q_1$ for different 
parameter settings. 
Let us point out that it is possible to modify the spectrum of the transformed problem further by employing 
a different choice for the constant $C$ in (\ref{u1ex2}). For example, if we choose $C=-5$, the 
discrete spectrum of the transformed problem (\ref{bvpx1t2}), (\ref{bvpx2t2}) will be given by
\begin{eqnarray}
(E_n) &=& \left\{\pm \sqrt{2~n+1+m^2}: n=0,1,2,3...\right\} \cup \left\{\pm \sqrt{4.1+m^2} \right\}, \nonumber
\end{eqnarray}
observe that the constant $C$ enters in all discrete spectral values.


\subsection{A third-order confluent SUSY transformation}
\begin{figure}[h]
\begin{center}
\epsfig{file=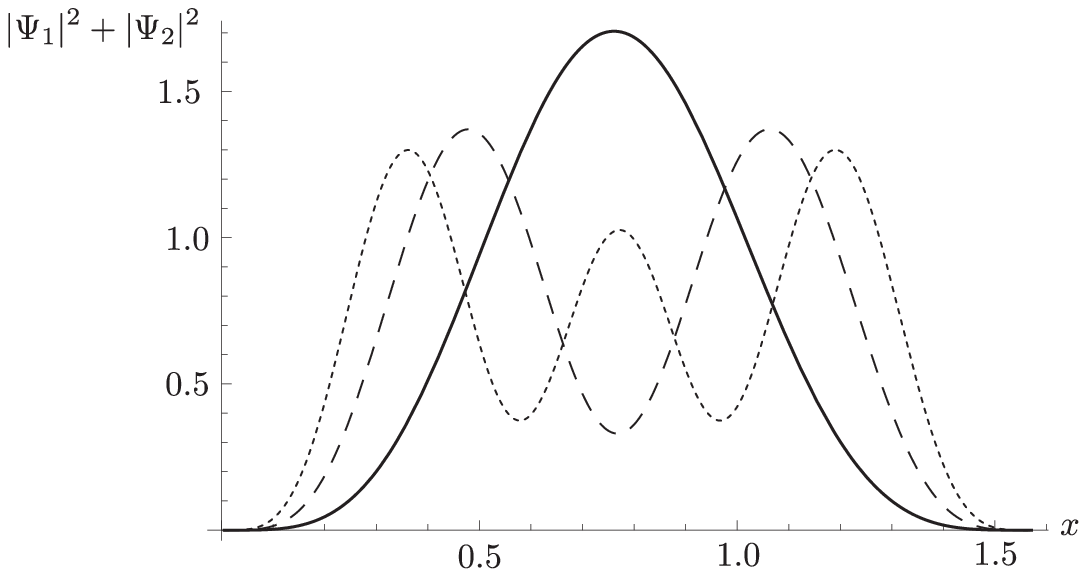,width=10cm}
\caption{Normalized probability densities $|\Psi_1|^2+|\Psi_2|^2$ for the solutions (\ref{psi1sol}), (\ref{psi2sol}) with 
the settings $m=1$, $n=0$ (solid curve), $n=1$ (dashed curve), and $n=2$ (dotted curve).} 
\label{prob3ini}
\end{center}
\end{figure}

\begin{figure}[h]
\begin{center}
\epsfig{file=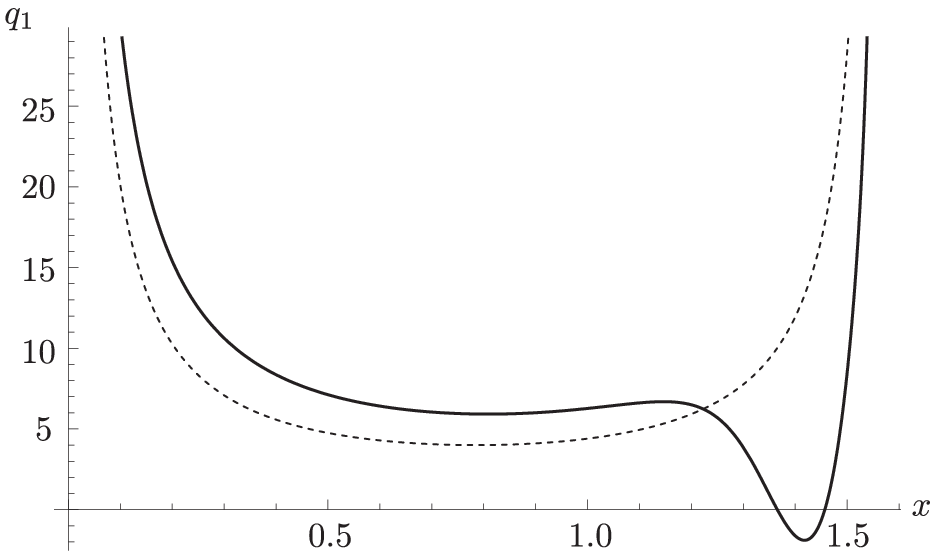,width=8cm}
\caption{The initial and transformed parametrizing functions $q_0$ (dotted curve) and $q_1$ (solid curve), as defined in (\ref{q0tan}) and 
(\ref{q1ex}).} 
\label{q0q1}
\end{center}
\end{figure}

\begin{figure}[h]
\begin{center}
\epsfig{file=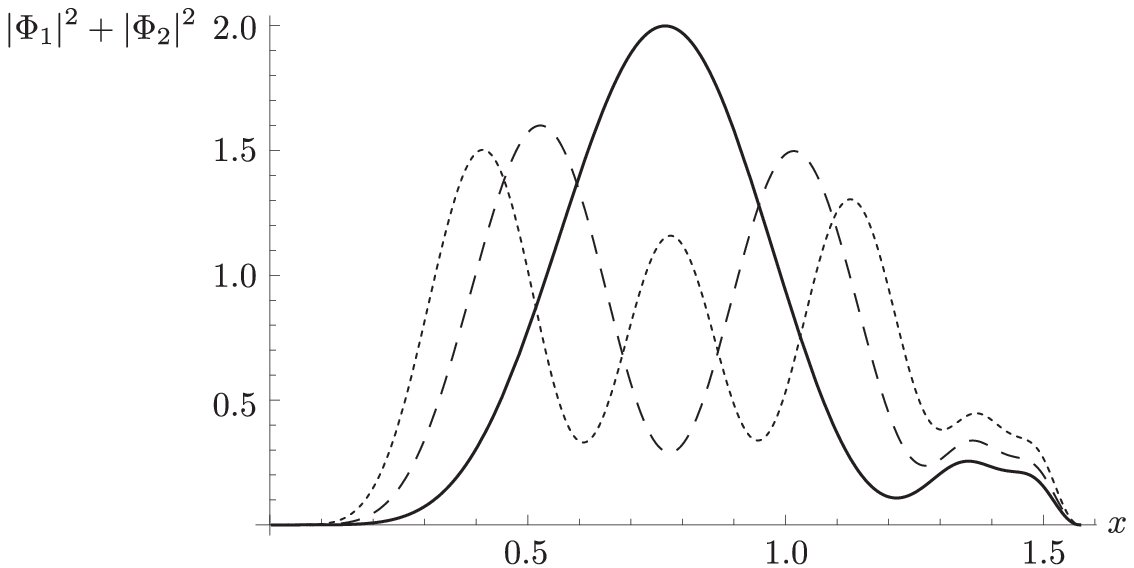,width=10cm}
\caption{Normalized probability densities $|\Phi_1|^2+|\Phi_2|^2$ for the solutions (\ref{phit}), (\ref{phi2tr}) with 
the settings $m=1$, $n=1$ (solid curve), $n=2$ (dashed curve), and $n=3$ (dotted curve).} 
\label{prob3trans}
\end{center}
\end{figure}

While the previous examples referred to the confluent SUSY algorithm of second order, let us now 
present a transformation of third order. To this end, we consider the folllowing boundary-value problem 
of Dirichlet type 
\begin{eqnarray}
i~\sigma_2~\Psi' + (V_0 - E)~\Psi &=& 0,~~~~x \in \left(0,\frac{\pi}{2} \right) \label{bvp1} \\[1ex]
\Psi(0) ~=~ \Psi\left( \frac{\pi}{2} \right) &=& \left( \begin{array}{l} 0 \\ 0 \end{array} \right), \label{bvp2}
\end{eqnarray}
where the pseudoscalar potential $V_0$ is given by
\begin{eqnarray}
V_0 &=& m~\sigma_3+2 \left[\tan(x)+\frac{1}{\tan(x)} \right]\sigma_1. \label{pseudoex}
\end{eqnarray}
Comparison of (\ref{pseudoex}) with the general form 
(\ref{pseudo}) shows that the parametrizing function $q_0$ reads
\begin{eqnarray}
q_0 &=& 2~\tan(x)+\frac{2}{\tan(x)}. \label{q0tan}
\end{eqnarray}
The boundary-value problem (\ref{bvp1})-(\ref{pseudoex}) admits an infinite discrete spectrum $(E_n)$ and 
a corresponding orthogonal set of solutions $(\Psi_n)$, where $n$ is a nonnegative integer. The spectral 
values are given by
\begin{eqnarray}
|E_n| &=& \sqrt{\left(2~n+5 \right)^2+m^2}, \label{ene}
\end{eqnarray}
while the associated solution spinor components $\Psi_1$, $\Psi_2$ can be expressed through 
hypergeometric functions as follows
\begin{eqnarray}
\Psi_1 &=& \cos^3(x)~\sin^2(x)~{}_2F_1\left[
-n, n+5,\frac{5}{2},\sin^2(x)
\right] \label{psi1sol} \\[1ex]
\Psi_2 &=& \frac{4~n~(n+5)~\cos^4(x)~\sin^3(x)}{5~m+5~\sqrt{(2~n+5)^2+m^2}}~
{}_2F_1\left[
-n+1,n+6,\frac{7}{2},\sin^2(x)
\right]+ \nonumber \\[1ex]
&+&
\frac{5~\cos^2(x)~\sin^3(x)}{m+\sqrt{(2~n+5)^2+m^2}}
~{}_2F_1\left[-n,n+5,\frac{5}{2},\sin^2(x)
\right]. \label{psi2sol}
\end{eqnarray}
Observe that the first argument of the hypergeometric functions is a nonpositive integer, such that they can 
be represented through Jacobi polynomials. For the sake of brevity, we omit to state the latter 
representation explicitly, and instead refer the reader to \cite{abram}. Figure \ref{prob3ini} shows 
normalized probability densities 
associated with the solutions (\ref{psi1sol}), (\ref{psi2sol}) for the first three values of $n$.
Now, in order to perform a 
third-order confluent SUSY transformation, we must find three auxiliary solutions $u_0,u_1,u_2$ that 
solve the Jordan chain (\ref{j1}), (\ref{j2}) for $N=3$. As a first step, we choose the function 
$u_0$ as follows:
\begin{eqnarray}
u_0 &=& \left(\Psi_1 \right)_{\mid n=0} ~=~ \cos^3(x)~\sin^2(x). \label{u0}
\end{eqnarray}
This choice determines the energy $\lambda$ in (\ref{j1}), (\ref{j2}) as $\lambda = 25$, which we obtained from (\ref{ene}) for $n=0$. Next, we use the integral 
representation (\ref{integral}) to find the remaining two auxiliary solutions $u_1$ and $u_2$. Substitution of 
(\ref{u0}) yields the result
\begin{eqnarray}
u_1 &=& \frac{4~x \left[\cos(8 x)+2~\cos(6 x)-2 ~\cos(4 x)-6~ \cos(2 x) \right]-\sin(6 x)+\sin(4 x)+11 ~\sin(2 x)}
{5120~\cos^2(x)~\sin(x)} \label{u1} \\[1ex]
u_2 &=& \frac{(51~x-2560)~\tan(x)}{256000~\cos(x)}+\frac{1}{768000} \left\{
6~\cos(x) \left[80~x^2 \cos(4x)-14~\cos(2x)-80~x^2+21\right]+ \right. \nonumber \\[1ex]
&+& \left. \frac{5~(-39~x+2560)}{\sin(x)}+
\frac{153}{\cos(x)}-8~(21~x-2560)~\left[\sin(3x)-2~\sin(x)\right]+ \right. \nonumber \\[1ex]
&+& \left. 16~(-3~x+1280)~\sin(5x)
\right\}, \label{u2}
\end{eqnarray}
note that we chose $\hat{u}=0$. Furthermore, we set all integration constants to zero except for the 
inner integral in $u_2$, where we picked the integration constant as $-1/20$. Now that the three auxiliary 
solutions have been determined, we are ready to perform 
the confluent SUSY transformation. To this end, we plug (\ref{psi1sol}), (\ref{u0})-(\ref{u2}) into (\ref{phi}):
\begin{eqnarray}
\Phi_1 &=& \frac{W_{u_0,u_1,u_2,\Psi_1}}{W_{u_0,u_1,u_2}}. \label{phit}
\end{eqnarray}
This function constitutes the first component of our transformed solution spinor $\Phi=(\Phi_1,\Phi_2)^T$. 
Since the resulting expression for $\Phi_1$ is very long, we omit to state it here in explicit form. Before we 
can determine the second component, the parametrizing function $q_1$ of the transformed potential 
in (\ref{diract}) must be constructed using (\ref{q1p}) or (\ref{q1}). We choose the simplest case (\ref{q1p}), 
given by
\begin{eqnarray}
q_1 &=& \frac{d}{dx}~ \log\left(\Phi_1{}_{\mid n = -\frac{5}{2}} \right). \label{q1ex}
\end{eqnarray}
Observe that choosing $C=0$ in \eqref{U1} and $n=-5/2$ results in $\epsilon=E^2-m^2=0$, that is, the function $\Phi_1$ is evaluated 
at zero energy $\epsilon$. Similar to $\Phi_1$, the function $q_1$ has a very long and involved form, such that 
we do not display it here. The same holds for the remaining component $\Phi_2$ of the transformed 
solution spinor $\Phi$ that is generated by substituting (\ref{phit}) and (\ref{q1ex}) into (\ref{phi2tr}). 
Figure \ref{q0q1} shows the parametrizing functions $q_0$ and $q_1$ of our initial and transformed 
potentials, respectively. Furthermore, normalized probability densities for our transformed solution spinor are 
visualized in figure \ref{prob3trans} for the first three values of $n$. Note that there is no solution for 
$n=0$ anymore, as the corresponding bound state was was deleted by our choice of $u_0$ in 
(\ref{u0}).


\section{Concluding remarks}
We have introduced the confluent SUSY formalism for the Dirac equation with a pseudoscalar potential. 
The present discussion can be easily extended to scalar potentials, as this requires only a 
reparametrization \cite{nieto}. The formalism was applied to three different interactions each one with different domain of definition, the first a Coulomb-like potential, then to a linear potential with a rational extension and finally to a trigonometric interaction. Some of the possible spectral manipulation that the formalism permits were illustrated. While in principle confluent transformations of arbitrary order can be 
performed using the formalism presented in this work, the calculations become very complicated, even 
when resorting to a symbolic calculator.

\section*{Acknowledgments}
ACA acknowledges Conacyt fellowship 207577.

\end{sloppypar}
\end{document}